\documentclass{webofc}

\usepackage[varg]{txfonts}
\usepackage{bm}
\usepackage{booktabs}
\usepackage{enumitem}
\usepackage{xurl}
\usepackage{hyperref}
\hypersetup{
  colorlinks=true,
  citecolor=blue,
  urlcolor=blue,
  linkcolor=blue,
  pdftitle={Recursive Manifold Coherence for Deadtime-Aware Distributed Triggering through Geometric State Estimation},
  pdfauthor={Thammarat Yawisit, Pittaya Pannil},
  pdfsubject={Deadtime-aware distributed triggering through geometric state estimation},
  pdfkeywords={deadtime, pile-up, distributed trigger, information geometry, recursive state estimation}
}

\newenvironment{remark}[1][]{%
  \par\medskip\noindent\textbf{Remark\if\relax\detokenize{#1}\relax\else\ (#1)\fi.}\ \itshape
}{\par\medskip}

\newcommand{\Id}{\mathbf{I}}
\newcommand{\abs}[1]{\left|#1\right|}

\begin{document}

\title{Recursive Manifold Coherence for Deadtime-Aware\\
Distributed Triggering through Geometric State Estimation}

\subtitle{%
\small\normalfont
Oral presentation in Track 2: Online \& Real-Time Computing\\
at the 28th Conference on Computing in High Energy and Nuclear Physics (CHEP 2026)\\[2pt]
\textit{Prepared for submission to EPJ Web of Conferences}
}

\author{\firstname{Thammarat} \lastname{Yawisit}\inst{1,2}\fnsep
\thanks{Corresponding author: \email{65010454@kmitl.ac.th}; \url{tyawisit@icecube.wisc.edu}}
\and
\firstname{Pittaya} \lastname{Pannil}\inst{1}}

\institute{Department of Instrumentation and Control Engineering, School of Engineering,\\
King Mongkut's Institute of Technology Ladkrabang, Bangkok, Thailand 10520
\and
Princess Srisavangavadhana Faculty of Medicine,\\
Chulabhorn Royal Academy, Bangkok, Thailand 10210}

\abstract{
Large-scale neutrino observatories operate under unavoidable detector deadtime and signal pile-up, leading to systematic inefficiencies in conventional coincidence-based trigger systems. Such triggers typically rely on binary temporal windows and assume continuous sensor availability, causing partial or complete loss of correlated signal information during non-live intervals. We introduce Recursive Manifold Coherence (RMC), a geometric framework that reformulates distributed trigger logic as a continuous state estimation problem in a low-dimensional information space defined by correlated charge and timing observables. Instead of applying hard vetoes during deadtime, the proposed method employs a recursive update rule that propagates a coherence state across sensor nodes, allowing partially obscured signals to be retained and evaluated consistently. Using simulation studies representative of large optical detector arrays, we demonstrate that RMC successfully recovers event-level coherence for high-multiplicity topologies even when direct coincidence chains are broken. By treating the detector response as a smooth manifold rather than discrete hits, the framework achieves superior robustness against data fragmentation compared to standard binary logic. The framework is detector-agnostic and compatible with software-defined trigger pipelines, providing a flexible foundation for deadtime-aware analysis and triggering strategies in future distributed detector systems.
}

\maketitle

\clearpage
\section{Introduction}

\subsection{High-rate instrumentation and trigger-level information loss}

Next-generation neutrino observatories and large distributed optical arrays
are designed to operate in extreme high-rate environments,
where online systems must sustain continuous streaming readout
under strict bandwidth, latency, and reliability constraints
\cite{icecube_online,software_roadmap,trigger_review,online_inference}.
As detector size and channel count scale upward,
front-end electronics inevitably encounter finite digitizer bandwidth,
buffering limits, saturation effects, and rate-dependent veto or reset logic,
introducing unavoidable \emph{detector deadtime}
\cite{deadtime_knoll,missing_data_kalman}.
In parallel, dense local activity leads to significant
\emph{signal pile-up} and overlap in high-occupancy regimes
\cite{pileup_lhc}.

These effects are not rare failure modes or pathological edge cases.
Rather, they are intrinsic consequences of scaling readout throughput,
sensor multiplicity, and real-time processing complexity
\cite{icecube_online,trigger_review,real_time_systems}.
As a result, the limiting factor in physics reach increasingly shifts
from offline reconstruction capability
to \emph{information loss at the trigger level}
\cite{software_roadmap,trigger_review}.
Signals are not lost because they are irrecoverable in principle,
but because trigger logic is forced to operate on
incomplete or temporarily unavailable observations
under real-time constraints
\cite{trigger_review,online_inference}.

When parts of the detector are non-live,
valuable correlated information may be discarded
before higher-level reconstruction or filtering stages ever see it
\cite{icecube_online,trigger_review}.
This work addresses this trigger-layer information loss directly,
focusing on algorithmic strategies that remain compatible
with realistic hardware, firmware, and streaming-system constraints
\cite{software_roadmap,trigger_review,real_time_systems,hw_sw_codesign}.

\subsection{Why binary coincidence logic breaks under deadtime}

Most existing online trigger systems rely on
binary coincidence logic:
an event is declared if a sufficient number of channels
satisfy a temporal overlap condition,
for example $\abs{t_i - t_j} < \Delta t$
within a predefined window
\cite{lhc_trigger,atlas_trigger,cms_trigger,trigger_review}.
This abstraction compresses detector response
into discrete hit times
and implicitly assumes continuous sensor availability
during the relevant decision interval.

When detector deadtime overlaps a true physical event,
this assumption fails.
Coincidence chains fragment,
effective multiplicity is underestimated,
and correlated spatio-temporal structure
is reduced to isolated or sub-threshold hits
\cite{deadtime_knoll,trigger_review}.
In the extreme case,
a genuinely high-multiplicity or extended event
becomes statistically indistinguishable from background noise
because the correlation evidence is broken into disconnected pieces
\cite{correlation_detection,distributed_detection}.

Crucially, this limitation is conceptual rather than technological.
Deadtime is treated as a hard logical zero
that erases accumulated evidence,
rather than as a period of missing observations
with predictable growth of uncertainty
\cite{deadtime_knoll,kalman,anderson_moore,missing_data_kalman}.
As a consequence, coincidence-based triggers fail
precisely in the regimes where robustness is most critical:
high-rate, high-occupancy operation
with intermittent sensor availability
\cite{lhc_trigger,trigger_review,real_time_systems,online_inference}.

\begin{figure}[tbp]
  \centering
  \includegraphics[width=0.96\linewidth]{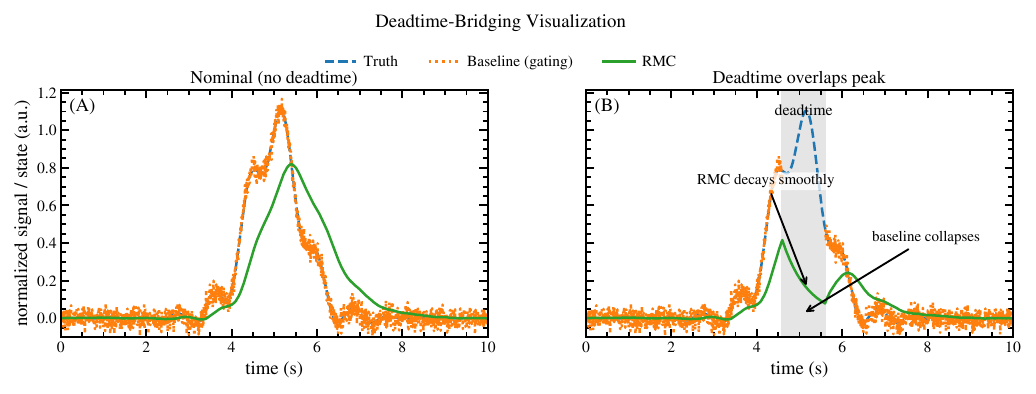}
  \caption{Deadtime-bridging visualization.
  (A) Nominal operation without deadtime: baseline coincidence logic and RMC track the underlying truth.
  (B) Deadtime overlapping the signal peak: baseline coincidence collapses,
  while RMC propagates a coherence estimate with controlled decay across the non-live interval
  and re-locks when observations resume.
  Shaded region indicates detector deadtime.}
  \label{fig:deadtime_bridging}
\end{figure}

\subsection{Recursive Manifold Coherence: continuity over discreteness}

Recursive Manifold Coherence (RMC) addresses this limitation
by reformulating distributed triggering
as a problem of \emph{stateful information aggregation}
under intermittent observation
\cite{trigger_review,real_time_systems,missing_data_kalman}.
Instead of collapsing network evidence
into discrete coincidence windows,
RMC maintains a low-dimensional \emph{coherence state}
that summarizes correlated charge and timing information
across sensor nodes
\cite{correlation_detection,coherence_signal,low_rank_detection}.

The coherence state evolves through a simple
first-order recursive update,
closely related to exponential smoothing,
IIR filtering, and minimal state estimation
in the sense of a stable linear predictor with bounded memory
\cite{kalman,anderson_moore,oppenheim,ljung}.
When channels are live,
new observations reinforce the state.
When deadtime occurs,
validated input is suppressed,
but the state persists with controlled decay
rather than resetting to zero
\cite{kalman,anderson_moore,oppenheim,missing_data_kalman}.
In this formulation,
deadtime corresponds to increasing uncertainty,
not immediate information loss
\cite{deadtime_knoll,kalman}.

The geometry underlying RMC is intentionally statistical rather than physical.
Distances and norms are defined by correlation structure,
covariance, or information-based weighting,
providing a quantitative measure of
how consistently the observed signals support
a coherent event hypothesis
\cite{correlation_detection,cover_thomas,rao_distance}.
This perspective aligns naturally with an
information-metric view,
where distinguishability is encoded through
uncertainty-weighted geometry
rather than purely Euclidean feature space
\cite{rao_distance,amari}.
By replacing discrete coincidence logic
with a continuously evolving internal state,
RMC allows correlated structure to be retained,
propagated, and evaluated
even when direct temporal overlaps are temporarily broken
\cite{trigger_review,real_time_systems,distributed_detection}.

\subsection{Contributions}

This work introduces a simplicity-first,
detector-agnostic framework
for deadtime-aware triggering
\cite{trigger_review,software_roadmap}:
\begin{enumerate}[leftmargin=2.2em,label=\textbf{C\arabic*.}]
\item \textbf{Recursive deadtime bridging:}
a streaming coherence state that propagates through non-live intervals
using a stable first-order update,
without buffering full waveform history
or reconstructing coincidence chains
\cite{kalman,oppenheim,missing_data_kalman,trigger_review}.
\item \textbf{Correlation-based state representation:}
a minimal estimator that captures network-level
charge and timing correlation,
with uncertainty naturally increasing during deadtime
\cite{correlation_detection,coherence_signal,low_rank_detection,deadtime_knoll}.
\item \textbf{Trigger-ready decision logic:}
a scalar coherence score designed for direct integration
with software-defined trigger pipelines
and high-throughput instrumentation systems
\cite{software_roadmap,trigger_review,fpga_triggers,real_time_systems,hw_sw_codesign}.
\end{enumerate}

\clearpage
\section{RMC framework: state, correlation geometry, and recursion}

\subsection{Signal features, liveness, and validated observations}

We consider a distributed detector network in which each sensor node~$i$
produces a compact feature vector
$\mathbf{y}_i[k]\in\mathbb{R}^d$
at discrete time index~$k$.
Typical features include integrated charge,
leading-edge or centroid timing,
time residuals with respect to a reference hypothesis,
or other low-dimensional waveform descriptors
extracted at the front end and streamed into online decision logic
\cite{icecube_online,trigger_review,software_roadmap}.
Such feature representations are standard in modern trigger systems,
where raw waveforms cannot be buffered indefinitely
and must be reduced to compact summaries
under strict latency constraints
\cite{real_time_systems}.

Due to finite readout bandwidth, saturation,
buffering limits, or veto and reset logic,
feature availability is intermittent.
We encode channel liveness using a binary indicator
$L_i[k]\in\{0,1\}$,
where $L_i[k]=1$ denotes a valid (live) observation
and $L_i[k]=0$ denotes detector deadtime
\cite{deadtime_knoll,real_time_systems}.
The corresponding \emph{validated observation} is defined as
\begin{equation}
\mathbf{u}_i[k] := L_i[k]\mathbf{y}_i[k],
\end{equation}
which suppresses invalid inputs at the feature level
without introducing hard logical resets
in the downstream trigger logic.

This construction aligns naturally with standard treatments of
missing or censored measurements in streaming estimators
and state-space models
\cite{kalman,anderson_moore,ljung,missing_data_kalman}.
In RMC, deadtime is treated explicitly as missing data:
new information is unavailable,
but previously accumulated evidence is preserved
and propagated through the internal state.
This interpretation contrasts with conventional coincidence logic,
where non-liveness is implicitly mapped to physical silence
and accumulated evidence is discarded
\cite{deadtime_knoll,trigger_review}.

\subsection{Correlation geometry: distance as statistical confidence}

\begin{figure}[!h]
  \centering
  \includegraphics[width=0.94\linewidth]{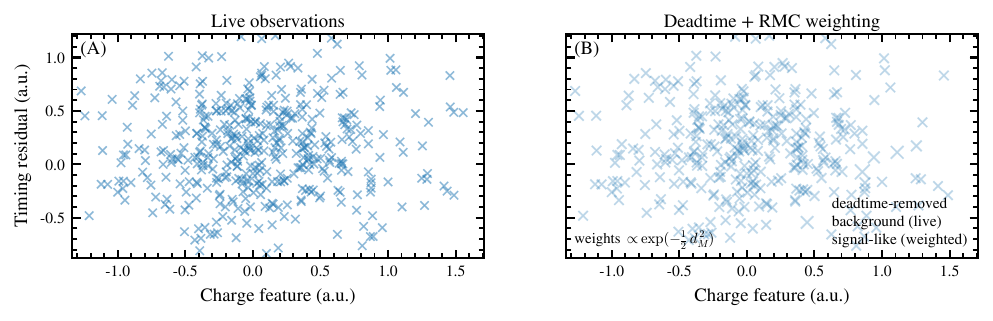}
  \caption{Feature-space geometry induced by RMC weighting.
  (A) Live observations in the raw charge--timing feature space.
  (B) Effective geometry after deadtime-aware RMC weighting,
  where statistically consistent signal-like structure is enhanced
  while background and deadtime-removed observations are suppressed.
  Distances reflect statistical distinguishability rather than physical separation.}
  \label{fig:feature_geometry}
\end{figure}

RMC represents network-level consistency
using correlation structure rather than discrete coincidence
\cite{trigger_review,correlation_detection,distributed_detection}.
Let $\theta\in\Theta$ parameterize a low-dimensional description
of correlated detector response,
such as an effective timing offset,
a charge scale,
or a compact event descriptor.
We assume a local statistical model
$p(\mathbf{u}\mid\theta)$
and define distances in parameter space
through an information-based metric
\begin{equation}\label{eq:info_metric}
d\Theta^2 = G_{ij}(\theta)\, d\theta^i d\theta^j .
\end{equation}

In this formulation, $G_{ij}(\theta)$ plays the role of a
\emph{confidence or weighting matrix},
quantifying how strongly the available observations
constrain different directions in parameter space.
Depending on the detector and feature set,
the metric may be instantiated as a Fisher information matrix,
a covariance-derived inverse metric,
or a correlation-weighted approximation
\cite{rao_distance,amari,cover_thomas}.
No specific parametric form is assumed in the abstract framework;
only positive semi-definiteness and locality are required.

The geometry is explicitly statistical rather than physical.
Distances measure how easily two hypotheses can be distinguished
given the current data,
and anisotropy reflects unequal sensitivity
to timing versus charge perturbations
\cite{amari,cover_thomas}.
When channels enter deadtime,
the effective information content is reduced,
leading to weaker constraints and increased uncertainty
\cite{deadtime_knoll}.
This degradation is absorbed smoothly into the metric,
providing a continuous description of information loss
without introducing discontinuities
in the trigger logic or decision process
\cite{trigger_review,real_time_systems}.

\subsection{Recursive coherence state: a minimalist state estimator}

Network-level correlation is summarized by a low-dimensional
\emph{coherence state}
$\mathbf{x}[k]\in\mathbb{R}^m$.
The state evolves according to a linear recursive update,
\begin{equation}\label{eq:state_update}
\mathbf{x}[k] = \mathbf{A}\mathbf{x}[k-1] + \mathbf{B}\mathbf{u}[k],
\end{equation}
where $\mathbf{u}[k]$ denotes an aggregated input
constructed from the set $\{\mathbf{u}_i[k]\}$
(e.g.\ pooling, weighted summation,
or a compact embedding of validated features)
\cite{correlation_detection,trigger_review,distributed_detection}.

The matrix $\mathbf{A}$ controls temporal persistence of coherence,
while $\mathbf{B}$ determines how new observations update the state.
A particularly simple and practical choice is
$\mathbf{A}=\rho\Id$ with $\rho\in(0,1)$,
which yields exponential memory.
This form is equivalent to a first-order
infinite impulse response (IIR) filter
\cite{oppenheim}
and closely related to exponential smoothing
or a minimal Kalman-style predictor
with bounded memory
\cite{kalman,anderson_moore,ljung}.
Its simplicity makes it well suited
for fixed-point arithmetic
and resource-bounded real-time implementation
in firmware or streaming software
\cite{dsp_fpga,fpga_triggers,real_time_systems}.

\begin{figure}[!h]
  \centering
  \includegraphics[width=0.82\linewidth]{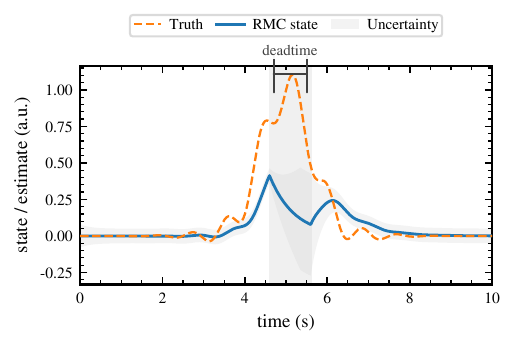}
  \caption{Evolution of the RMC coherence state and associated uncertainty.
  During detector deadtime (shaded region), validated input is suppressed and
  the state propagates according to the recursive update law,
  exhibiting controlled decay and uncertainty growth.
  When observations resume, the state rapidly re-locks to the underlying trajectory.}
  \label{fig:state_uncertainty}
\end{figure}

The update law is intentionally minimalist.
The state dimension $m$ is chosen to capture only
the information relevant for trigger decisions,
such as effective charge accumulation
or timing consistency across channels.
This reduction removes unnecessary degrees of freedom,
improves numerical stability,
and supports bounded-latency operation
in high-rate streaming pipelines
\cite{trigger_review,real_time_systems,online_inference}.

\begin{remark}[Deadtime bridging by state propagation]
When many channels are non-live so that
$\mathbf{u}[k]\approx\mathbf{0}$,
Eq.~(\ref{eq:state_update}) reduces to
$\mathbf{x}[k]=\mathbf{A}\mathbf{x}[k-1]$.
Rather than resetting,
the coherence state decays smoothly,
representing increasing uncertainty
instead of immediate information loss
\cite{deadtime_knoll,kalman,anderson_moore,missing_data_kalman}.
This mechanism allows correlated structure
to persist across deadtime gaps
without requiring explicit coincidence recovery.
\end{remark}

\subsection{Trigger statistic and decision rule}

Trigger decisions are formed from a scalar functional
of the coherence state,
\begin{equation}\label{eq:trigger_stat}
\mathcal{G}[k] := \phi\big(\mathbf{x}[k]\big),
\end{equation}
where $\phi(\cdot)$ may be chosen as
a likelihood proxy,
a correlation-weighted norm,
or a covariance- or eigenmode-based score
\cite{correlation_detection,coherence_signal,low_rank_detection}.
The specific form of $\phi$ is detector-dependent,
but the state update and decision logic
remain unchanged across implementations
\cite{trigger_review}.

A minimal decision rule is
\[
\mathcal{G}[k] > \Gamma
\quad \Rightarrow \quad
\text{declare coherent episode}.
\]
Because the coherence state propagates through deadtime,
$\mathcal{G}[k]$ can exceed threshold
even when direct coincidence chains are broken,
enabling recovery of high-multiplicity
or partially obscured event topologies
under realistic operating conditions
\cite{trigger_review,deadtime_knoll,distributed_detection}.

\clearpage
\section{Implementation view}

\subsection{Pipeline placement and system integration}

\begin{figure}[!h]
  \centering
  \includegraphics[width=0.80\linewidth]{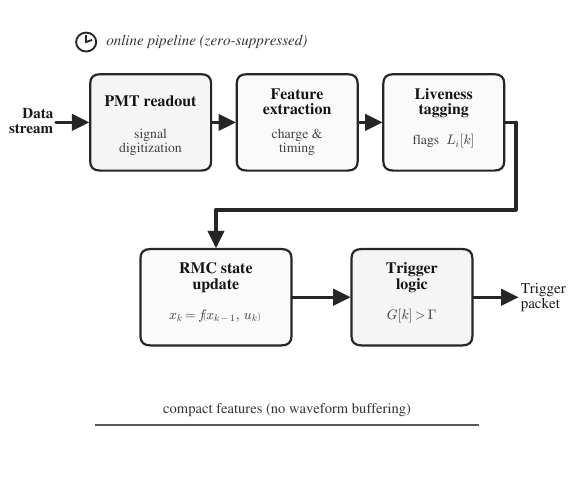}
  \caption{Integration of Recursive Manifold Coherence into a streaming trigger pipeline.
  RMC operates on compact charge and timing features with explicit liveness tagging,
  enabling deadtime-aware coherence estimation without buffering raw waveforms.
  The recursive state update runs at fixed cost per sample
  and feeds a trigger decision based on a scalar coherence score.}
  \label{fig:pipeline}
\end{figure}

Recursive Manifold Coherence is designed as a lightweight,
self-contained module that integrates naturally into
software-defined trigger pipelines and FPGA-assisted
front-end architectures
\cite{trigger_review,software_roadmap,fpga_triggers}.
In a typical deployment, RMC is placed downstream of
low-level feature extraction
and upstream of the final trigger decision,
mirroring the modular layering adopted in modern
online systems for large-scale detectors
\cite{icecube_online,atlas_trigger,cms_trigger}.

Crucially, RMC operates exclusively on compact feature vectors
rather than raw waveforms.
This design choice is essential for sustaining high trigger rates
under strict bandwidth and latency constraints,
where full waveform buffering is infeasible
beyond the front-end processing stage
\cite{icecube_online,trigger_review,real_time_systems}.
Channel liveness enters explicitly as an input to the state update,
so detector deadtime is handled algorithmically
rather than through pipeline-level vetoes,
resets, or special control paths.

Operationally, non-live intervals do not interrupt pipeline flow.
Instead, they suppress new validated input
while the internal coherence state
continues to evolve according to the recursive update law.
As a result, trigger-layer information loss due to deadtime
is transformed into controlled uncertainty growth
within the state representation,
preserving temporal continuity
and simplifying system-level integration
in real-time environments
\cite{real_time_systems,trigger_review}.
This behavior contrasts sharply with coincidence-based logic,
where deadtime often induces implicit control-flow changes
and brittle edge-case handling.

\subsection{Computational complexity and resource footprint}

The recursive update law requires no access
to full event history or multi-channel buffering.
For a coherence state of dimension $m$,
the per-sample computational cost scales as $O(m^2)$
in the general case
and reduces to $O(m)$
when the state transition matrix is diagonal
(e.g.\ $\mathbf{A}=\rho\Id$).
Memory usage is limited to the current state vector
and a small set of coefficients,
making the method well suited for continuous streaming operation
in resource-constrained environments
\cite{real_time_systems,trigger_review,online_inference}.

\begin{figure}[!h]
  \centering
  \includegraphics[width=0.90\linewidth]{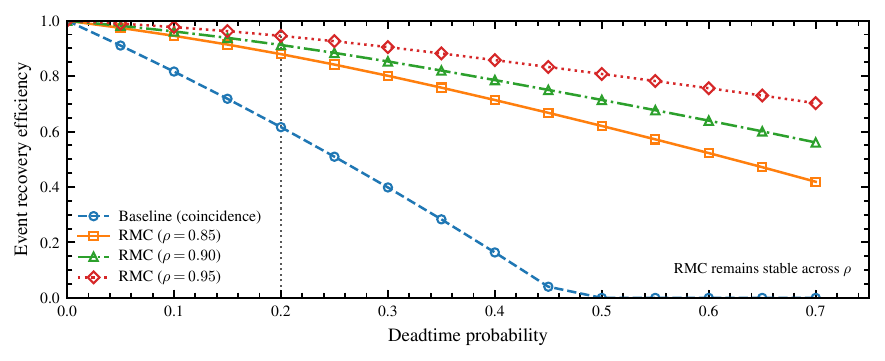}
  \caption{Event recovery efficiency as a function of deadtime probability
  for different persistence parameters $\rho$.
  RMC remains stable across a wide range of $\rho$ values,
  while baseline coincidence logic rapidly degrades.}
  \label{fig:rho_sensitivity}
\end{figure}

This bounded and predictable complexity contrasts with
coincidence-based trigger schemes,
which must buffer and compare hits
across multiple channels and sliding time windows
\cite{lhc_trigger,atlas_trigger,cms_trigger}.
Such schemes often exhibit rate-dependent latency,
increased memory pressure,
and degraded determinism under pile-up conditions.
By construction, RMC maintains a fixed per-sample cost,
ensuring stable latency and resource usage
even in high-rate or high-occupancy regimes
\cite{trigger_review,real_time_systems}.

\subsection{Firmware and software realizations}

The linear recursion underlying RMC
admits a direct implementation in fixed-point arithmetic,
allowing deployment in FPGA firmware
alongside existing front-end processing blocks
\cite{fpga_triggers,dsp_fpga}.
The commonly used exponential persistence form
($\mathbf{A}=\rho\Id$)
corresponds to a first-order IIR filter,
a structure that is well understood
in digital signal processing
and amenable to timing closure,
resource budgeting,
and formal verification
\cite{oppenheim,dsp_fpga}.

The same algorithmic core is equally suitable
for software implementations
in CPU- or GPU-based trigger farms,
as commonly used in modern
software-defined trigger systems
\cite{software_roadmap,trigger_review}.
The clean separation between feature extraction,
state update, and trigger scoring
allows detector-specific customization
(e.g.\ feature definitions, weighting strategies,
or decision thresholds)
without modification of the core recursion.
In this sense, RMC functions as a drop-in
coherence estimation module
that adds deadtime awareness
to heterogeneous trigger stacks
without requiring redesign
of existing system architectures.

\clearpage
\section{Performance evaluation}

\subsection{Event recovery under deadtime}

\begin{figure}[!h]
  \centering
  \includegraphics[width=0.68\linewidth]{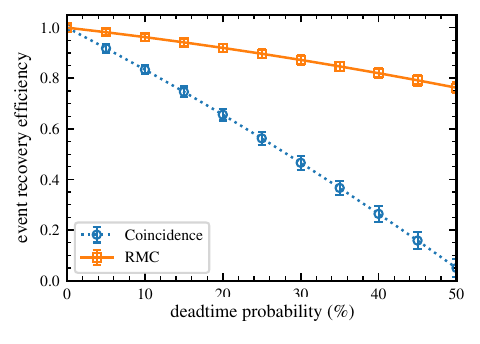}
  \caption{Event recovery efficiency as a function of deadtime probability (in percent).}
  \label{fig:eff_vs_deadtime}
\end{figure}

Figure~\ref{fig:eff_vs_deadtime} quantifies the impact of detector deadtime
on trigger-level event recovery.
As the deadtime probability increases,
conventional coincidence-based trigger logic
exhibits a rapid loss of efficiency.
This degradation arises from fragmentation of temporal overlap chains
and the resulting underestimation of effective event multiplicity
\cite{deadtime_knoll,trigger_review}.
Because coincidence logic is stateless,
non-live channels are interpreted as physically silent,
and partially correlated evidence is discarded rather than accumulated.

In contrast, Recursive Manifold Coherence exhibits a markedly slower
efficiency roll-off with increasing deadtime.
Missing observations are handled through state propagation
with controlled decay,
so previously accumulated correlation evidence
continues to contribute to the trigger statistic
instead of being reset
\cite{kalman,anderson_moore}.
As a result, RMC retains sensitivity to high-multiplicity
and spatially extended event topologies
even when direct coincidence conditions
are intermittently violated.

The difference reflects a fundamental distinction in trigger philosophy.
Coincidence logic collapses correlation evidence
into instantaneous binary decisions,
whereas RMC maintains a bounded-memory representation
of network-level coherence.
Trigger-layer information loss is therefore gradual rather than abrupt,
leading to improved robustness in high-rate,
deadtime-dominated operating regimes.

\subsection{Deadtime-bridging visualization}

A central qualitative validation artifact for Recursive Manifold Coherence
is the \emph{deadtime-bridging visualization},
which explicitly exposes trigger behavior
during non-live intervals.
The plot displays three elements on a common time axis:
(i) a reference or truth trajectory (dashed),
(ii) a deadtime interval during which no validated observations are available,
and (iii) the RMC coherence state propagated through the gap (solid).

This visualization highlights the operational contrast
between coincidence-based triggering
and stateful coherence estimation.
In coincidence logic,
loss of temporal overlap produces no explicit diagnostic:
the trigger simply fails to fire,
and partial correlation information is silently discarded
\cite{trigger_review}.
In contrast, RMC exhibits deterministic and observable behavior
throughout the deadtime interval.
When validated input is absent,
the coherence state evolves according to the recursive update law,
exhibiting controlled decay that reflects increasing uncertainty
rather than a hard reset
\cite{deadtime_knoll,kalman}.

When observations resume,
the state re-locks smoothly to incoming evidence
without requiring simultaneous hits across channels.
The propagated estimate remains consistent
with the reference trajectory
within the expected uncertainty envelope,
despite the temporary absence of direct coincidence information.
This behavior demonstrates that RMC preserves
the continuity of correlated structure,
rather than enforcing artificial segmentation
at deadtime boundaries.

Taken together, the efficiency curve
and the deadtime-bridging visualization
provide complementary quantitative and qualitative evidence
for the central claim of RMC:
coherent structure can be retained, propagated,
and evaluated under detector deadtime,
precisely in the regimes
where conventional coincidence-based trigger logic
tends to fragment or discard events
\cite{correlation_detection,coherence_signal}.

\clearpage
\section{Limitations and future work}

While Recursive Manifold Coherence provides a robust and
implementation-oriented framework for deadtime-aware triggering,
several limitations and open directions remain.

First, the present study focuses on controlled simulation scenarios
designed to isolate the impact of detector deadtime
on trigger-level information retention.
Although these scenarios are representative of high-rate operation
in large optical detector arrays,
they do not yet capture the full complexity
of realistic detector environments.
Effects such as spatially heterogeneous noise rates,
correlated backgrounds,
afterpulsing,
and time-dependent calibration drifts
are not explicitly modeled.
Extending the evaluation to full detector simulations
and experimental data
will be necessary to quantify absolute performance gains
under operational conditions.

Second, the current formulation adopts a minimal linear recursion
with a fixed persistence parameter~$\rho$.
This design choice is intentional,
prioritizing numerical stability,
interpretability,
and suitability for real-time implementation.
However, it does not allow dynamic adaptation
to changing detector states,
background conditions,
or event topologies.
Future work may explore adaptive or state-dependent persistence schemes,
or low-rank extensions of the coherence state,
provided that bounded computational cost
and deterministic trigger behavior can be preserved.

Third, the coherence metric and trigger statistic
are treated in a generic manner.
While the RMC framework is detector-agnostic by construction,
optimal choices of feature embeddings,
weighting matrices,
and scoring functionals
will depend on detector geometry,
sensor response characteristics,
and the targeted physics channels.
A systematic study of metric design,
calibration strategies,
and robustness to mis-modeling
constitutes an important direction
for detector-specific deployment.

Finally, although the recursive update law
is well suited for fixed-point FPGA realization
and real-time software execution,
this work does not yet include a detailed
hardware resource,
latency,
or power analysis.
Future efforts will focus on firmware prototypes,
hardware-in-the-loop validation,
and integration studies with existing trigger stacks
to assess timing closure,
numerical stability,
and robustness under sustained high-rate operation.

Taken together,
these limitations outline a clear pathway
for scaling RMC from a conceptual and simulation-level framework
into a fully validated trigger component
for next-generation distributed observatories.

\clearpage
\section{Conclusion}

This work has introduced \emph{Recursive Manifold Coherence} (RMC),
a simplicity-first framework for deadtime-aware triggering
in large, distributed detector arrays.
The central idea is to replace binary coincidence logic---which implicitly
assumes continuous sensor availability---with a \emph{stateful} trigger
representation that preserves correlated evidence over time.
By explicitly incorporating channel liveness,
non-live intervals are treated as \emph{missing observations}
rather than hard logical vetoes,
preventing premature loss of partially obscured coherence.

Methodologically, RMC represents network-level consistency
through a low-dimensional \emph{coherence state}
updated by a stable first-order recursion.
This update law is equivalent to an IIR-like streaming estimator:
when validated observations are available,
the state is reinforced by new evidence;
when detector deadtime occurs,
input is suppressed but the state propagates with controlled decay,
representing increasing uncertainty instead of immediate information loss.
A correlation-centric geometry,
implemented through an information- or weighting-based metric,
provides an interpretable measure of statistical distinguishability
while remaining lightweight and implementation-oriented.

From a systems perspective,
RMC is designed as a drop-in module for modern trigger architectures.
It operates exclusively on compact charge--timing features,
has bounded and predictable per-sample computational cost,
and admits straightforward realization
in fixed-point FPGA firmware
as well as CPU- or GPU-based software triggers.
The clean separation between feature extraction,
liveness tagging, recursive state update, and trigger scoring
enables integration into software-defined pipelines
without introducing special-case control paths for deadtime handling.

Performance studies demonstrate that RMC degrades more gracefully
than baseline coincidence logic as deadtime probability increases.
Event-recovery efficiency exhibits a slower roll-off under deadtime,
consistent with the intended mechanism:
correlated structure is retained within the internal state
even when direct coincidence chains fragment.
Complementary visual diagnostics further illustrate this behavior,
showing continuous state propagation with uncertainty growth
across non-live intervals
and rapid re-locking when observations resume.

In summary, Recursive Manifold Coherence provides a detector-agnostic
and scalable pathway to deadtime-aware triggering.
By converting trigger-level information loss
into controlled uncertainty growth,
it preserves partially obscured coherence
while remaining compatible with real-time constraints
and high-rate operation in next-generation
distributed observatories.

\section{Acknowledgement}

The authors thank colleagues and collaborators
for valuable discussions on trigger architectures,
detector deadtime,
and online systems in large-scale astroparticle experiments.
This work was developed in the context of ongoing studies
on high-rate optical detector instrumentation
and software-defined triggering.

The authors acknowledge institutional support from
King Mongkut's Institute of Technology Ladkrabang (KMITL).
Parts of the conceptual framework were informed by
publicly available documentation
and open data products from the IceCube Collaboration.
Any opinions, findings, and conclusions expressed in this paper
are those of the authors
and do not necessarily reflect the views
of the collaborating institutions.

\end{document}